# Ferroelectricity in artificial bicolor oxide superlattices**

By *Sung Seok A. Seo, Jun Hee Lee, Ho Nyung Lee,\* Matthew F. Chisholm, Woo Seok Choi, Dong Jik Kim, Ji Young Jo, Hanchul Kim, Jaejun Yu, and Tae Won Noh\**

[*] Prof. T. W. Noh, Dr. S. S. A. Seo, W. S. Choi, Dr. D. J. Kim, J. Y. Jo
Research Center for Oxide Electronics (ReCOE)
Frontier Physics Research Division (FPRD)
Department of Physics and Astronomy, Seoul National University, Seoul 151-747 (Korea)
E-mail: twnoh@snu.ac.kr
Prof. J. Yu, J. H. Lee
Center for Strongly Correlated Materials Research (CSCMR)
Frontier Physics Research Division (FPRD)
Department of Physics and Astronomy, Seoul National University, Seoul 151-747 (Korea)
Dr. H. N. Lee, Dr. M. F. Chisholm
Materials Science and Technology Division,
Oak Ridge National Laboratory, Oak Ridge, TN 37831 (USA)
E-mail: hnlee@ornl.gov
Dr. H. Kim
Division of Chemical Metrology and Materials Evaluation
Korea Research Institute of Standards and Science (KRISS)
Daejeon 305-600 (Korea)

[**] We thank J. G. Yoon and T. K. Song for useful advice and D. Vanderbilt for valuable discussion. This work was financially supported by Creative Research Initiatives (Functionally Integrated Oxide Heterostructure), CSCMR, and the National R&D Project for Nano Science and Technology of MOST/KOSEF. The experiments at PLS were supported by MOST and POSTECH. H.N.L. and M.F.C. gratefully acknowledge the experimental assistance of C. M. Rouleau and H. M. Christen, and the support by the Division of Materials Sciences and Engineering, Office of Basic Energy Sciences, U.S. Department of Energy, under contract DE-AC05-00OR22725 with Oak Ridge National Laboratory, managed and operated by UT-Battelle, LLC.

Compositional superlattices, which consist of a periodic sequence of ultrathin layers of different compounds,[1] have attracted a lot of attention for decades, because these artificial structures can exhibit intriguing physical phenomena that are not observable in bulk materials. The most typical example is a semiconductor superlattice, where compositional modulation using precise thickness control can create an artificially periodic electronic potential on a length scale shorter than the mean free path of carriers. These artificial structures have provided us with a wealth of novel physical phenomena, some of which have led to important technological applications, including cascade lasers.[2] When sufficiently thin layers are used, even heterostructures with lattice-mismatched semiconductors can be used to grow a superlattice with essentially no misfit dislocations.[3]

Recently, advanced techniques in complex oxide synthesis made also similar breakthroughs, providing new opportunities to search for new intriguing physical properties.[4, 5] There have also been advances in modeling and computational tools to explore physical properties of ferroelectric oxides.[6, 7] For instance, theoretical studies have predicted microscopic phenomena that occur in the interface between ferroelectric and non-ferroelectric oxide layers,[8] and near a metallic electrode.[9] The theoretical insights have also stimulated experimental investigations into related fundamental issues, such as the nanometer-scale heterostructures of ferroelectric oxides,[10] the strain control of ferroelectricity,[5, 11, 12] and the ferroelectric critical thickness.[13]

In this paper we report on the growth and properties of high quality bicolor oxide superlattices, composed of two perovskites out of $BaTiO_3$ (BTO), $CaTiO_3$ (CTO), and $SrTiO_3$ (STO). We grew the bicolor superlattices on $SrRuO_3$/STO (001) single-crystalline substrates by pulsed laser deposition. Before growing superlatttices, conducting $SrRuO_3$ films were grown as a bottom electrode on single-stepped, $TiO_2$-terminated STO substrates.[14] Figures 1a shows a schematic diagram of sample structure. As one can see in Fig. 1a, short-period superlattices with two alternating layers have important advantages that facilitate the process of fabrication, since it experiences less strain relaxation than that with a larger periodicity. Moreover, alternating compressive and tensile strain states can cancel out the opposite-directional mechanical tension. Therefore, the maximum effect can be found in (BTO)$n$(CTO)$n$ superlattices, because BTO has the largest in-plane lattice constants and CTO has the smallest among the three perovskites. Note that BTO is tetragonal ($a = b = 3.995$ Å, $c = 4.0335$ Å) and CTO is pseudo-cubic ($a = 3.822$ Å) at room temperature. For (CTO)$n$(STO)$n$ and (BTO)$n$(STO)$n$ superlattices, the STO layers will make lattice relaxation much less likely occur. Therefore, we fabricated (BTO)$n$(CTO)$n$, (CTO)$n$(STO)$n$, and (BTO)$n$(STO)$n$ superlattices, all of which have one or two unit cell periodicity ($n \leq 2$).

In bulk crystal forms, BTO is ferroelectric at room temperature with its Curie temperature of ~120 $^o$C,[15] but CTO and STO are paraelectric. Using very thin layers of BTO, CTO, and STO, we fabricated highly strained superlattices probably without significant misfit dislocations as demonstrated by a cross-sectional Z-contrast scanning transmission electron microscopy image of (BTO)2(STO)2 superlattices (Fig. 1b). Here, (BTO)2(STO)2 means that the schematic superlattice is composed of alternating BTO and STO layers of two-unit-cell thickness. For constructing such atomic-scale superlattices, we controlled the deposition process by monitoring the specular-spot intensity of reflection high-energy electron diffraction (RHEED), as shown in Fig. 1c. The well defined oscillation of RHEED specular spot intensity confirms clearly that the layers are grown layer-by-layer with atomistic precision. Figure 1d shows typical x-ray diffraction patterns of (BTO)$n$(CTO)$n$ superlattices with different periodicity $n$, and the clear satellite peaks demonstrate that they are grown with the artificial periodicities along the c-axis with nearly perfect crystallinity. The total thickness of all superlattices is 100 or 200 nm.

It is also worthy to note that the (BTO)$n$(CTO)$n$ superlattices are unique structures from the viewpoint of chemical composition. By contrast with the $Ba_{1-x}Sr_xTiO_3$ and $Sr_{1-x}Ca_xTiO_3$ solid solutions, the $Ba_{1-x}Ca_xTiO_3$ solid solution does not naturally exist in bulk forms in the composition region near $x=0.5$ due to the solubility limit. The ionic sizes of Ba and Ca are considerably different to each other, so the phase segregation is thermodynamically more favorable. Namely, in the composition region near $x=0.5$, the bulk material should be a mixture of the two phases of lightly Ca doped BTO and lightly Ba doped CTO.[16] Therefore, the (BTO)$n$(CTO)$n$ superlattices should be an intriguing type of artificial structure, where the Ba and Ca ions are mixed homogeneously at the macroscopic scale.

In order to check the ferroelectricity of the superlattices, we employed piezoelectric force microscopy (PFM), a powerful tool to control/observe ferroelectric domain structures.[17] Figure 2a shows the domain pattern, which is written by PFM tip, and observed in a (BTO)2(CTO)2 superlattice. The amplitude and the phase stand for the amount and the direction of polarization, respectively. A square domain with 1.0×1.0 μm$^2$ in size was written with +10.5V dc-bias voltage ($V_{dc}$), while an inner square domain of 0.5×0.5 μm$^2$ was oppositely written with $V_{dc}=-10.5V$ in order to switch the polarization in the superlattice sample. Small amplitude signal in the inner square domain is due to the asymmetry of top and bottom surface in the superlattice. Figure 2b shows the piezoelectric response (amplitude and phase) as a function of $V_{dc}$ with as a small ac voltage ($V_{ac}\sin\omega t$) applied to the bottom electrode in the contact mode[18]. The written domain in Fig. 2a and the observed 180$^o$ shift of phase with $V_{dc}$ in Fig. 2b reveal that the (BTO)2(CTO)2 superlattice might have ferroelectricity.

Seo *et al.*

Figure 2c displays the polarization *vs.* electric field (*P–E*) hysteresis loops of the superlattices at room temperature. The (BTO)2(CTO)2, (BTO)1(CTO)1, and (BTO)1(STO)1 superlattices clearly show ferroelectric hysteresis loops, while (CTO)1(STO)1 reveals a paraelectric behavior. The room temperature value of $P_r$ for (BTO)2(CTO)2 and (BTO)1(CTO)1 is ~8.3 µC/cm$^2$. Our experimental observation of constant $P_r$ regardless of *n* seems to be consistent with a simple electrostatic model[5, 8] in the (BTO)*n*(CTO)*n* superlattices (*n*≤2):

$$P_r = P_{FE}/(1+(t_{PE}/t_{FE})/(\varepsilon_{PE}/\varepsilon_{FE})). \qquad (1)$$

Here, $P_{FE}$ is the polarization in a ferroelectric layer, and $t_{PE}/t_{FE}$ and $\varepsilon_{PE}/\varepsilon_{FE}$ are the thickness and the dielectric permittivity ratio, respectively, between the paraelectric (CTO) and ferroelectric (BTO) layers. Note that this $P_r$ value of the superlattices is enhanced by a factor of 4, compared to that of pseudo-binary ceramics of *(0.5)*BTO–*(0.5)*CTO, *i.e.* ~2 µC/cm$^2$.[19] Ferroelectric hysteresis loops can be affected by the leakage current or the trapped charges at interfaces. However, since PFM (shown in Fig. 2a and b) measures the mechanical (piezoelectric) property in ferroelectrics, *P-E* hysteresis loops in Fig. 2c are consequences neither of the leakage currents nor of the trapped charges. By comparison, (BTO)1(STO)1 has a $P_r$ of ~2.8 µC/cm$^2$, which is less than that of (BTO)1(CTO)1. The smaller polarization value of (BTO)1(STO)1 is somewhat unexpected by considering the fact that STO is in closer vicinity to the ferroelectric instability than CTO, because STO is an incipient ferroelectric material. Therefore, we performed a short-time pulse measurement[20] in order to accurately measure the polarization by eliminating any significant reduction by short-time polarization relaxation. The measured polarization values are shown as solid dots in Fig. 2a. $P_s$ of the (BTO)1(CTO)1 and (BTO)1(STO)1 superlattices are found to be ~14 and ~10 µC/cm$^2$, respectively.

In order to understand the mechanism of observed ferroelectricity in the unit cell period superlattices, we performed first-principles calculations. The in-plane lattice constants of the superlattices were fixed to the theoretically determined value of cubic SrTiO$_3$ (*a*=3.866 Å), and the ions were allowed to move along the *c*-axis during structural optimization. Initially, as shown in Fig. 3a, the ionic positions were relaxed, maintaining the inversion symmetry of the system. The relaxed geometry with this constraint clearly reveals a local anti-ferroic distortion. Then, as shown in Fig. 3b, the ions were allowed to be fully relaxed by removing the constraint of the inversion symmetry. The ionic displacements in the fully relaxed structure break the inversion symmetry and yield a net polarization in each unit cell.

By interpolating between the structures of Figure 3a and b, we calculated the total energy curves for the (BTO)1(CTO)1, (BTO)1(STO)1, and (CTO)1(STO)1 superlattices as a function of polarization. As shown in Fig. 3c, the (BTO)1(CTO)1 and (BTO)1(STO)1 superlattices represent double-well potential energy surfaces with double minima at ±22.10 and ±20.15 $\mu C/cm^2$, respectively, confirming the ferroelectric ground state. On the other hand, the (CTO)1(STO)1 superlattice shows a potential energy surface with a single minimum at zero polarization, typical paraelectrics. These theoretical $P_s$ values are somewhat larger than the experimentally measured values. Such discrepancies might come from structural imperfections in our real superlattices, different temperatures between the experiment (room temperature) and the computation (zero temperature), and/or the theoretical imprecision in calculating the lattice constant values. However, it is still noteworthy that the $P_s$ value of (BTO)1(CTO)1 is larger than that of (BTO)1(STO)1, which is qualitatively consistent with our experimental observation.

The ionic motions particularly in ferroelectrics are very important, because they break the inversion symmetry and provide the net dipole moments for displacive ferroelectrics. The arrows in Fig. 3b show the calculated displacements of the ions in the (BTO)1(CTO)1 superlattice, whose values are also listed in Table I. In order to further visualize the ionic displacements that occur in the ferroelectric state, we plotted the change in valence-electron density between the ferroelectric (Fig. 3b) and non-ferroelectric (Fig. 3a) states by setting Ba ions as the origin. The left and right-hand images in Fig. 4a illustrate cross-sectional views of the valence-electron density changes within the $Ti_2O_4$ and $CaBaO_2$ planes, respectively. Here, the red and green colors indicate the increase of valence-electron density in the forward direction of ionic motions and the decrease in the backward direction, respectively. For comparison, similar calculations were also performed for a fully strained tetragonal BTO film on a STO substrate, as shown in Fig. 4b. Note that the ferroelectric polarization of the BTO film can be described by the opposite ionic displacements of titanium and oxygen ions.

From the comparison between Figs. 4a and 4b, it is clear that the Ca ion motion is quite unique in the (BTO)1(CTO)1 superlattice: namely, the displacement of the Ca ion is large enough to induce a polarization in the CaO layer. The Ca ions move in a direction opposite to the motion of the oxygen ions, making the atomic corrugation in the CaO layer as large as 0.078 Å, which is comparable with that of 0.077 Å in the BaO layer. Although its Born effective charges are small, the Ca ion motions can still significantly contribute to the spontaneous polarization of the (BTO)1(CTO)1 superlattice. This is evidenced by the computed ionic polarizations shown in Table I: the contribution of Ca ions is comparable with those of other ions. If the CaO layer does not develop such a distortion, a large energy cost is required to build up discontinuous polarization charges between the originally non-ferroelectric CTO and the ferroelectric BTO layers. To minimize this energy cost, the ferroelectricity in BTO should be spread into the originally non-ferroelectric layers by the electrostatic coupling, resulting in a bending distortion in the interfacial CaO layer. Similar effects at the interface were also discussed by Neaton and Rabe[8], and Tenne et al.[21], for the (BTO)*m*(STO)*n* superlattices. In addition, the structural distortion at the hetero-interface can be also interpreted as a consequence of an elastic interaction, originating from the large difference in ionic size between Ba and Ca. Therefore, we think that these hetero-interfacial coupling play a crucial role in sustaining overall a high polarization in the (BTO)1(CTO)1 superlattice, where none of the $TiO_2$ layers have the same chemical environments as those in pure BTO or CTO.

In conclusion, we successfully fabricated A-site modulated, atomic-scale bicolor superlattices. The artificially grown superlattices are structurally unique and have a macroscopically homogeneous phase, which is not feasible to recreate in bulk form. By artificial structuring, it is found that the polarization of such superlattices can be highly increased as compared to pseudo-binary ceramics with the same overall composition. Such strong enhancement in superlattice is attributed to newly-developed ionic motions of A-site cations at the hetero-interfaces due to the interfacial coupling of electrostatic and elastic interactions, which cannot be found in single phase materials.

## *Experimental*

Single-crystalline STO(001) substrates with a miscut angle of <0.1° (CrysTec, Berlin) were used. The STO substrates were etched in a buffered $NH_4F$-HF(buffered oxide etch (BOE)):$H_2O$=1:10 solution (pH≈4.5) and annealed in air at temperatures between 1100 and 1200°C. The superlattices and conducting SrRuO$_3$ layers were grown by pulsed laser deposition (PLD), employing a KrF laser (λ=248 nm) and a multi-target carousel. Sintered ceramic targets SrRuO$_3$, CTO, and BTO and a single crystal STO target were used. The SrRuO$_3$ layers were deposited at a substrate temperature of 700°C in 100 mTorr O$_2$. Since the thickness of the SRO layer was quite thin (typically, less than 1000 Å), its in-plane lattice constants were found to be the same as those of STO substrates, *i.e.* 3.905 Å. Bicolor superlattices were grown at a substrate temperature of 700°C in 10 mTorr O$_2$ with monitoring a specular spot intensity of RHEED. More details on the PLD synthesis of superlattices can be found elsewhere.[5] Atomic resolution Z-contrast images were obtained using an aberration-corrected VG Microscopes HB603 scanning transmission electron microscope operated at 300 kV.

Piezoelectric force microscopy was measured with a conducting tip (Ti-Pt coated NSC18 cantilever, Mikromeasch) as moving top electrode at the surface of superlattice by XE-100 atomic force microscope (PSIA) and a SR830 lock-in amplifier (Stanford research). The RMS value of $V_{ac}$ was 0.5 V with 17 kHz.

*P-E* hysteresis loops were obtained by measuring the voltage of a sensing capacitor in Sawyer-Tower circuit with Yokogawa DL7100 digital oscilloscope and Yokogawa FG300 function generator or integrating of the displacement current, applying a triangular pulse of 100 kHz to the capacitor with top electrode (Au or Pt). In order to obtain a precise value of the spontaneous polarization, we applied pulse measurement which measured the non-switching polarization with write- and read-pulses using an arbitrary waveform generator.[20] The widths of write and read pulse were 5 μs, and the interval time was 1 μs, which is short enough to neglect the polarization relaxation. The non-switching current profiles obtained in a read pulse were acquired by measuring the voltage difference across the input impedance (50 Ω) of a digital oscilloscope connected in series to the superlattice capacitors. The non-switching polarization values were obtained by integrating the corresponding current profiles. Because the width of displacement current peak was within 100 ns, we integrated the current only for 100 ns to eliminate the leakage current. Thus, we obtained the non-switching polarizations in each electric field of different applied voltages of read-pulse with the negligible leakage current and polarization relaxation.

In the first principles calculation, we used local density approximation (LDA) and projector augmented wave (PAW) method with the Vienna *Ab initio* Simulation Package (VASP).[22] Cut-off energy for the plane-waves was 700 eV and the *k*-points were sampled from the 6×6×3 grid. Force relaxation was done until 0.001 eV/Å. The Born effective charges ($Z^*$) were calculated by the Berry-phase polarization[7]. The in-plane lattice constants of the superlattices were fixed to theoretical lattice constants of cubic STO (a=3.866 Å). The spontaneous polarization is obtained by using the formula for the unit-cell volume Ω:

$$P = \frac{1}{\Omega}\sum_i \int_{l_i^{zero}}^{l_i^{up}} Z_i^*(l)\,dl \approx \frac{1}{\Omega}\sum_i Z_i^*\,dl_i \quad ,$$

where $Z_i^*$ is the averaged Born effective charges, *l* is the positions of each ions at *i*. $l^{up}$ and $l^{zero}$ are indicating the states without and with the inversion symmetry of the system, respectively.

*Figures and Captions*

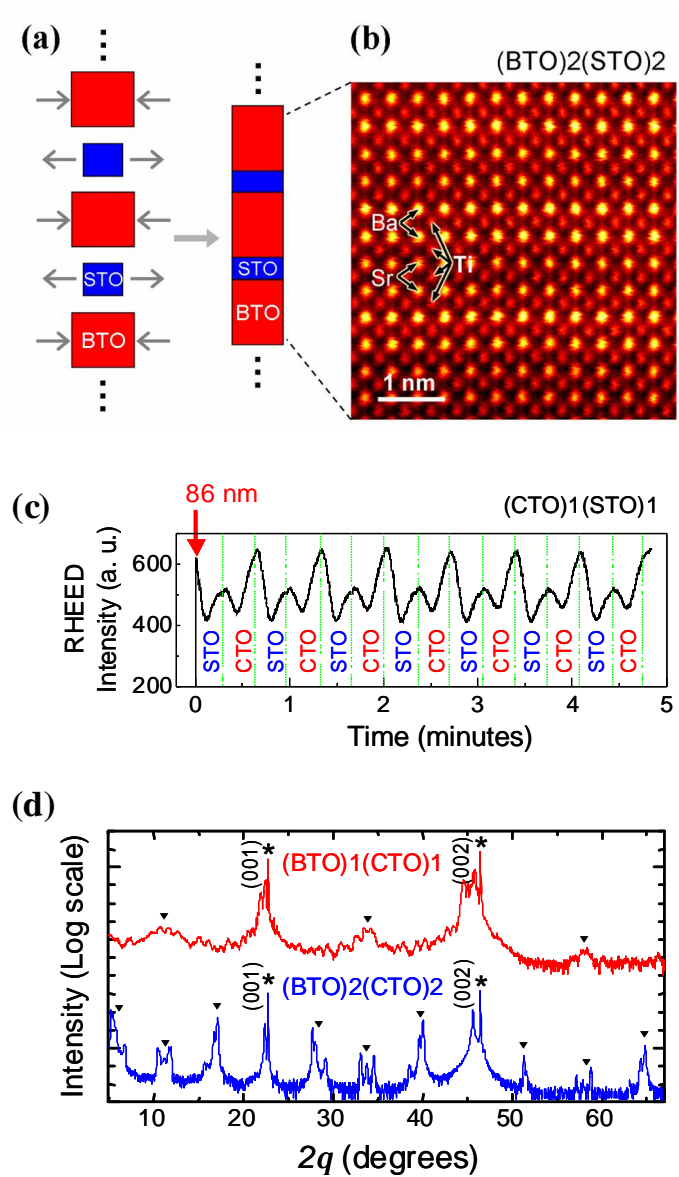

**Figure 1** Structures of bicolor superlattices. (a) Schematic diagram of a $(BaTiO_3)n(SrTiO_3)n$ bicolor superlattice. (b) High resolution Z-contrast image of a $(BaTiO_3)2(SrTiO_3)2$ superlattice. (c) Reflection high energy electron diffraction intensity oscillations during the growth of $(CaTiO_3)1(SrTiO_3)1$ one-unit-cell period superlattice. The red arrow indicates the time when the thickness of superlattice reaches around 86 nm (220 unit-cells of $CaTiO_3$ and $SrTiO_3$) in the middle

of deposition. (d) X-ray $\theta$-$2\theta$ diffraction patterns of (BaTiO$_3$)$n$(CaTiO$_3$)$n$ superlattices. Superlattice peaks due to artificial periodicity are marked by the filled triangles (▼). The asterisks (*) indicate peaks from SrTiO$_3$ substrates.

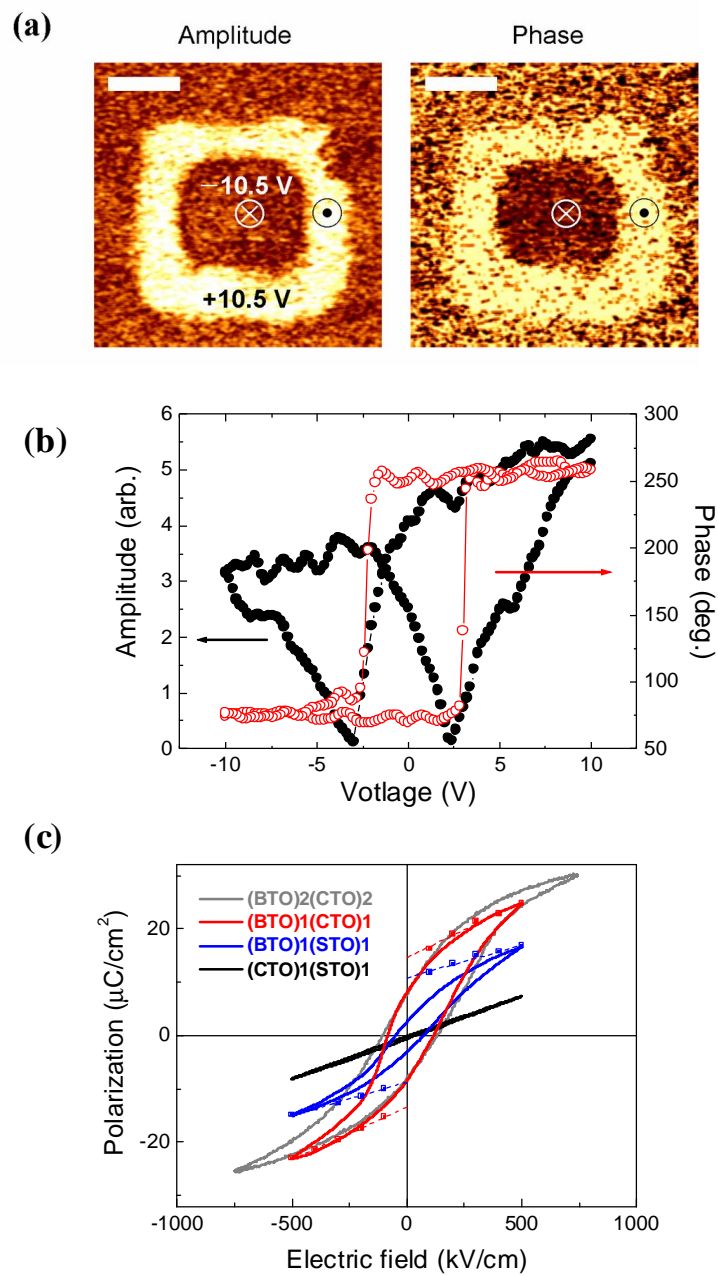

Seo *et al.*

**Figure 2** Piezoelectric force microscopy (PFM) results and ferroelectric *P-E* hysteresis loops of bicolor superlattices. (a) PFM images of $(BaTiO_3)2(CaTiO_3)2$ with the scale-bar of 500 nm. The outer square and the inner square are positive- and negative-poled, respectively. (b) Piezoelectric response switching as a function of dc-bias onto the ferroelectric superlattice. The signal shows hysteresis behavior clearly. (c) Room temperature *P-E* curves for short period superlattices. The observed remanent polarizations for $(BaTiO_3)1(CaTiO_3)1$ and $(BaTiO_3)1(SrTiO_3)1$ superlattices are ~8.3 and ~2.7 $\mu C/cm^2$, respectively, while the $(CaTiO_3)1(SrTiO_3)1$ superlattice demonstrates a paraelectric behavior. The solid circles indicate results of the short-time pulse measurements. The spontaneous polarizations $P_s$ are estimated to be ~14 and ~10 $\mu C/cm^2$ for $(BaTiO_3)1(CaTiO_3)1$ and $(BaTiO_3)1(SrTiO_3)1$, respectively.

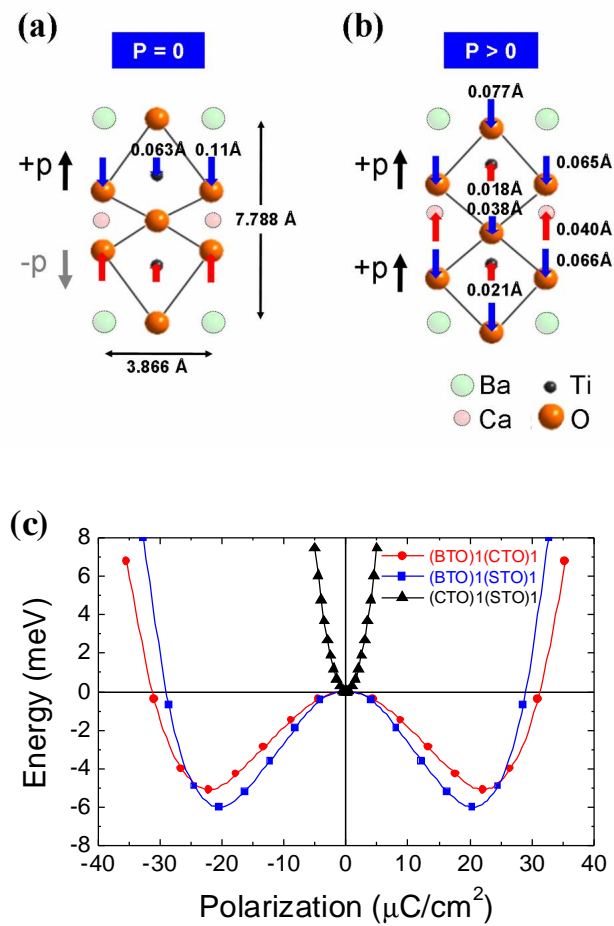

**Figure 3** Theoretical results on ferroelectricity of one-unit-cell period bicolor superlattices. Schematic diagrams of the calculated ground state for $(BaTiO_3)1(CaTiO_3)1$, (a) when the inversion symmetry is enforced during the relaxation of atomic positions, and (b) when the inversion symmetry constraint is removed during the relaxation. This provides the ferroelectric ground state. (c) Calculated total energy surfaces as a function of polarization. The two energy-minima, *i.e.* ferroelectric ground states, can be clearly seen from both $(BaTiO_3)1(CaTiO_3)1$ (red filled circles) and $(BaTiO_3)1(SrTiO_3)1$ (blue filled squares), and the spontaneous polarizations are ~22.10 and ~20.15 $\mu C/cm^2$, respectively. On the other hand, the $(CaTiO_3)1(SrTiO_3)1$ superlattice (black filled triangles) is found to develop a single-well energy surface, indicating that it should be in a paraelectric state.

Seo *et al.*

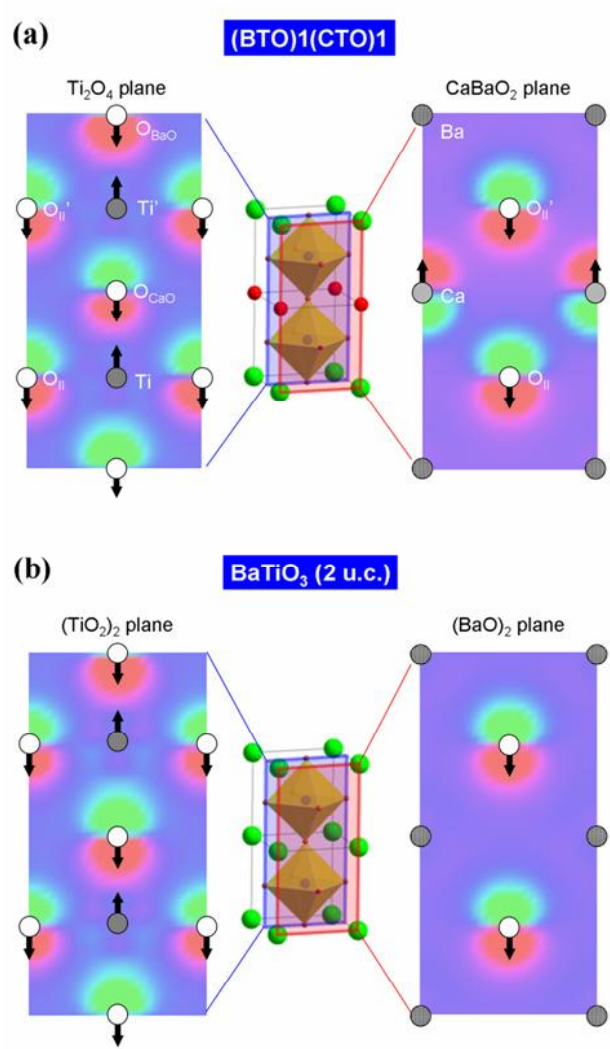

**Figure 4** Changes in the valence-electron density (ρ). ρ(P=+P$_r$)-ρ(P=0) from the first-principles calculations, as obtained by setting the Ba ions as the fixed origin: (a) the (BaTiO$_3$)1(CaTiO$_3$)1 superlattice, and (b) BaTiO$_3$, with the in-plane lattice parameters fixed to those of SrTiO$_3$. The ionic displacement pattern, which is responsible for the ferroelectric polarization, is clearly visualized. Note that the displacement of the Ca ions is rather large in the CaO layer, which should also contribute to the ferroelectric polarization of (BaTiO$_3$)1(CaTiO$_3$)1.

**Table I.** Born effective charges $Z^*_i$ and ionic displacements $l_i$ for each atom $i$ in the (BaTiO$_3$)1(CaTiO$_3$)1 and (BaTiO$_3$)1(SrTiO$_3$)1 superlattices. All values were obtained by first-principles calculations. $P_i$ corresponds to $Z^*_i \cdot l_i$, representing the polarization contribution from each atom $i$.

|  | Ba | Ca | Ti | Ti' | O$_{CaO}$ | O$_{BaO}$ | O$_{II}$ | O$_{II}$' |
|---|---|---|---|---|---|---|---|---|
| $Z^*_i$ (e) | 2.649 | 2.413 | 7.105 | 7.105 | -6.111 | -5.501 | -2.204 | -2.204 |
| $l_i$ (Å) | Fixed | 0.040 | 0.021 | 0.018 | -0.038 | -0.077 | -0.066 | -0.065 |
| $P_i$ (μC/cm$^2$) | – | 1.339 | 2.044 | 1.711 | 3.205 | 5.793 | 2.014 | 1.975 |

|  | Ba | Sr | Ti | Ti' | O$_{SrO}$ | O$_{BaO}$ | O$_{II}$ | O$_{II}$' |
|---|---|---|---|---|---|---|---|---|
| $Z^*_i$ (e) | 2.733 | 2.588 | 7.176 | 7.176 | -5.852 | -5.222 | -2.101 | -2.101 |
| $l_i$ (Å) | Fixed | 0.012 | 0.035 | 0.032 | -0.039 | -0.065 | -0.048 | -0.049 |
| $P_i$ (μC/cm$^2$) | – | 0.409 | 3.429 | 3.136 | 3.083 | 4.586 | 1.367 | 1.387 |

References

[1] K. Ploog, G. H. Dohler, *Advances in Physics* **1983**, *32*, 285.
[2] P. L. Gourley, *Nature* **1994**, *371*, 571.
[3] D. L. Smith, C. Mailhiot, *Reviews of Modern Physics* **1990**, *62*, 173.
[4] A. Ohtomo, D. A. Muller, J. L. Grazul, H. Y. Hwang, *Nature* **2002**, *419*, 378.
[5] H. N. Lee, H. M. Christen, M. F. Chisholm, C. M. Rouleau, D. H. Lowndes, *Nature* **2005**, *433*, 395.
[6] R. E. Cohen, *Nature* **1992**, *358*, 136.
[7] R. D. Kingsmith, D. Vanderbilt, *Physical Review B* **1993**, *47*, 1651.
[8] J. B. Neaton, K. M. Rabe, *Applied Physics Letters* **2003**, *82*, 1586.
[9] J. Junquera, P. Ghosez, *Nature* **2003**, *422*, 506.
[10] T. Shimuta, O. Nakagawara, T. Makino, S. Arai, H. Tabata, T. Kawai, *Journal of Applied Physics* **2002**, *91*, 2290.
[11] K. J. Choi, M. Biegalski, Y. L. Li, A. Sharan, J. Schubert, R. Uecker, P. Reiche, Y. B. Chen, X. Q. Pan, V. Gopalan, L. Q. Chen, D. G. Schlom, C. B. Eom, *Science* **2004**, *306*, 1005.
[12] J. H. Haeni, P. Irvin, W. Chang, R. Uecker, P. Reiche, Y. L. Li, S. Choudhury, W. Tian, M. E. Hawley, B. Craigo, A. K. Tagantsev, X. Q. Pan, S. K. Streiffer, L. Q. Chen, S. W. Kirchoefer, J. Levy, D. G. Schlom, *Nature* **2004**, *430*, 758.
[13] D. D. Fong, G. B. Stephenson, S. K. Streiffer, J. A. Eastman, O. Auciello, P. H. Fuoss, C. Thompson, *Science* **2004**, *304*, 1650.
[14] H. N. Lee, H. M. Christen, M. F. Chisholm, C. M. Rouleau, D. H. Lowndes, *Applied Physics Letters* **2004**, *84*, 4107.
[15] W. J. Merz, *Physical Review* **1953**, *91*, 513.
[16] T. Mitsui, W. B. Westphal, *Physical Review* **1961**, *124*, 1354.
[17] A. Gruverman, S. V. Kalinin, *Journal of Materials Science* **2006**, *41*, 107.
[18] S. Hong, J. Woo, H. Shin, J. U. Jeon, Y. E. Pak, E. L. Colla, N. Setter, E. Kim, K. No, *Journal of Applied Physics* **2001**, *89*, 1377.
[19] X. S. Wang, H. Yamada, C. N. Xu, *Applied Physics Letters* **2005**, *86*, 022905.
[20] D. J. Kim, J. Y. Jo, Y. S. Kim, Y. J. Chang, J. S. Lee, J. G. Yoon, T. K. Song, T. W. Noh, *Physical Review Letters* **2005**, *95*, 237602.
[21] D. A. Tenne, A. Bruchhausen, N. D. Lanzillotti-Kimura, A. Fainstein, R. S. Katiyar, A. Cantarero, A. Soukiassian, V. Vaithyanathan, J. H. Haeni, W. Tian, D. G. Schlom, K. J. Choi, D. M. Kim, C. B. Eom, H. P. Sun, X. Q. Pan, Y. L. Li, L. Q. Chen, Q. X. Jia, S. M. Nakhmanson, K. M. Rabe, X. X. Xi, *Science* **2006**, *313*, 1614.
[22] G. Kresse, J. Hafner, *Physical Review B* **1993**, *47*, R558.